\documentclass[a4paper]{article}

\usepackage{INTERSPEECH2022}
\usepackage{cite}
\usepackage{amssymb}
\usepackage{amsfonts}
\newcommand{\tabincell}[2]{\begin{tabular}
{@{}#1@{}}#2\end{tabular}}
\usepackage{multirow}
\usepackage{makecell}
\usepackage{color}

\title{Two-pass Decoding and Cross-adaptation Based System Combination of End-to-end Conformer and Hybrid TDNN ASR Systems}
\name{Mingyu Cui$^1$, Jiajun Deng$^1$, Shoukang Hu$^1$, Xurong Xie$^2$, Tianzi Wang$^1$, Shujie Hu$^1$, Mengzhe Geng$^1$, Boyang Xue$^1$, Xunying Liu$^1$, Helen Meng$^1$}
\address{
  $^1$The Chinese University of Hong Kong, Hong Kong SAR, China\\
  $^2$Institute of Software, Chinese Academy of Sciences, China}
\email{\footnotesize{\{mycui,jjdeng,skhu,tzwang,sjhu,mzgeng,byxue,xyliu,hmmeng\}@se.cuhk.edu.hk, xurong@iscas.ac.cn}}

\begin{document}

\bstctlcite{IEEEexample:BSTcontrol}

\maketitle
\begin{abstract}
  Fundamental modelling differences between hybrid and end-to-end (E2E) automatic speech recognition (ASR) systems create large diversity and complementarity among them. This paper investigates multi-pass rescoring and cross adaptation based system combination approaches for hybrid TDNN and Conformer E2E ASR systems. In multi-pass rescoring, state-of-the-art hybrid LF-MMI trained CNN-TDNN system featuring speed perturbation, SpecAugment and Bayesian learning hidden unit contributions (LHUC) speaker adaptation was used to produce initial N-best outputs before being rescored by the speaker adapted Conformer system using a 2-way cross system score interpolation. In cross adaptation, the hybrid CNN-TDNN system was adapted to the 1-best output of the Conformer system or vice versa. Experiments on the 300-hour Switchboard corpus suggest that the combined systems derived using either of the two system combination approaches outperformed the individual systems. The best combined system obtained using multi-pass rescoring produced statistically significant word error rate (WER) reductions of 2.5\% to 3.9\% absolute (22.5\% to 28.9\% relative) over the stand alone Conformer system on the NIST {\bf Hub5’00}, {\bf Rt03} and {\bf Rt02} evaluation data. 
\end{abstract}

\begin{figure*}[t]
\vspace{-1.5cm}
\setlength{\belowcaptionskip}{-0cm}
  \centering
  \includegraphics[width=\linewidth]{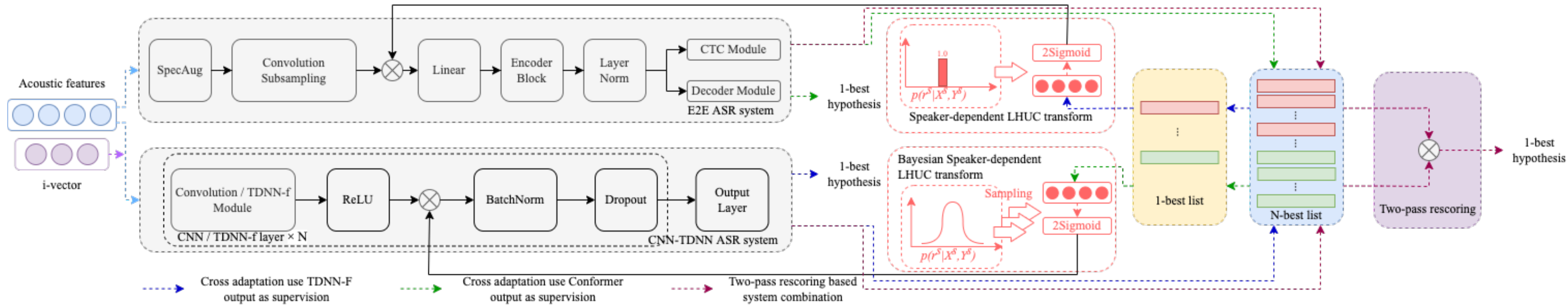}
  \caption{Two-pass rescoring (red dotted line) and cross adaptation (blue and green dotted lines) based system combination between  Bayesian LHUC adapted hybrid CNN-TDNN (grey, left bottom) and LHUC adapted E2E Conformer (grey, left top) ASR systems. }
  \label{fig:architecture}
\vspace{-0.6cm}
\end{figure*}

\vspace{-0.2cm}
\section{Introduction}
\vspace{-0.1cm}

Performance of automatic speech recognition (ASR) systems have been significantly advanced across a wide range of application domains in recent years. Such success was brought by a flux of deep neural networks (DNNs) based system architectures that can be categorized into two main types of design. The first is the traditional hybrid Hidden Markov model DNN (HMM-DNN) architecture\cite{vesely2013sequence, povey2018semi, abdel2012applying, medsker2001recurrent, graves2013speech, sak2014long} represented by time-delay neural networks (TDNNs)\cite{peddinti2015time, povey2016purely, povey2018semi}. The second is the end-to-end (E2E) paradigm\cite{chan2016listen, graves2012sequence, vaswani2017attention, dong2018speech, gulati2020conformer, guo2021recent}. Among these, the recently intruduced convolution-augmented Transformer (Conformer)\cite{guo2021recent} based E2E ASR systems have been successfully applied to a wide range of task domains. Compared with the earlier forms of Transformer models\cite{dong2018speech, karita2019comparative}, Conformer benefits from a combined use of self-attention and convolution structures designed to learn both the longer range and local contexts.

Fundamental modelling differences exist between hybrid and E2E ASR systems. The traditional hybrid HMM-DNN architecture is based on a structured, modular design with separate dedicated modelling components employed to represent acoustic, phonetic and lexical knowledge before they are combined to produce the most likely word recognition during recognition. A notable feature of hybrid ASR systems inherited from earlier HMM based systems is the recognition step is often performed in a frame synchronous manner over confusable phonetic units that are time aligned against input acoustic features. In contrast, a single neural network is used by E2E systems to directly convert input sequence of frames into output labels, thus simplifying the overall system design. The latent alignment between the input frames and output labels are often learned using attention mechanisms, for example, in attention encoder-decoder based models such as Conformer. The recognition stage is performed in a label synchronous fashion over confusable output tokens, while considering both the latent encoder representation computed over multiple input frames and the previous output token history. 

One general solution to exploit diversity and complementarity among ASR systems is to use system combination techniques that have been widely studied in the context of tradition HMM based systems\cite{woodland2004superears, schwartz2004speech, lei2009development, chu20102009, liu2013language,lamel2011improved}. To this end, three major categories of techniques can be used: a) cross system recognition outputs rescoring using multi-pass decoding\cite{woodland19951994, hain2005automatic, hain2003automatic}; b) cross system adaptation which requires the acoustic or language models of one system are adapted to the recognition outputs of another\cite{woodland19951994, woodland2004superears, peskin1999improvements, schwartz2004speech, prasad20052004, liu2013language}, thus allowing one system to learn the distinct and complementary properties from another system; and c) hypothesis level combination using ROVER\cite{fiscus1997post} or confusion network combination (CNC)\cite{evermann2000posterior}.

The large modelling differences between hybrid and E2E ASR systems and have also in recent years drawn increasing interest in developing suitable combination approaches to exploit their complementarity within the speech community\cite{watanabe2017hybrid, sainath2019two, evermann2000posterior, li2019integrating}. However, these prior researches were mainly conducted in the context of combining non-Conformer based E2E architectures such as CTC, LAS and RNN transducers\cite{watanabe2017hybrid, sainath2019two, evermann2000posterior, li2019integrating}.To the best of our knowledge, there has been no system combination approaches investigated for the combination between state-of-the-art hybrid TDNN and E2E Conformer based ASR systems. 

To this end, this paper presents the first use of multi-pass rescoring and cross adaptation based system combination approaches for TDNN and Conformer based ASR systems. In the two-pass rescoring based approach, state-of-the-art hybrid LF-MMI trained CNN-TDNN system\cite{peddinti2015time, povey2016purely} featuring speed perturbation, SpecAugment\cite{park2019specaugment} and Bayesian learning hidden unit contributions (LHUC) speaker adaptation\cite{xie2019blhuc, wong2020combination, xie2021bayesian} was used to produce initial N-best outputs before being rescored using LHUC speaker adapted Conformer\cite{deng2022confidence} using a 2-way cross system score interpolation. In cross adaptation based system combination, the hybrid CNN-TDNN system was cross adapted to the 1-best output of the Conformer system or vice versa. Experiments on the 300-hour Switchboard corpus suggest that the combined systems derived using either of the two forms of system combination approaches outperformed the individual systems. The best combined system obtained using multi-pass rescoring produced statistically significant word error rate (WER) reductions of 2.5\% to 3.9\% absolute (22.5\% to 28.9\% relative) over the stand alone Conformer system on the NIST Hub5’00, Rt03 and Rt02 evaluation sets. The efficacy of the proposed system combination techniques is further demonstrated in a comparison against the state-of-the-art performance obtained on the same task using the most recent hybrid and end-to-end systems reported in the literature.
\vspace{-0.2cm}
\section{E2E Conformer System}
\vspace{-0.1cm}
The Conformer E2E ASR system and the associated LHUC Speaker Adaptation method are presented in this section.
\vspace{-0.2cm}
\subsection{Conformer}
\vspace{-0.1cm}
The baseline convolution-augmented Transformer (Conformer) E2E ASR model considered in this paper follows the architecture proposed in \cite{gulati2020conformer, guo2021recent}. It consists of a Conformer encoder and a Transformer decoder. The Conformer encoder is based on a multi-blocked stacked architecture. Each encoder block includes the following components in turn: a position wise feed-forward (FFN) module, a multi-head self- attention (MHSA) module, a convolution (CONV) module and a nal FFN module at the end. Among these, the CONV module further consists of in turn: a 1-D pointwise convolution layer, a gated linear units (GLU) activation \cite{dauphin2017language}, a second 1-D point- wise convolution layer followed a 1-D depth-wise convolution layer, a Swish activation and a nal 1-D pointwise convolution layer. Layer normalization (LN) and residual connections are also applied to all encoder blocks. An example Conformer architecture is shown in Figure 1 (grey box, middle), 
A conformer architecture is illustrated in Figure~\ref{fig:architecture}(grey box, top). In Conformer training, the following multitask criteria by using interpolation of the CTC and attention cost is adopted \cite{xie2019blhuc}, 
\vspace{-0.1cm}
\begin{equation}
  \mathcal{L} = (1 - \lambda)\mathcal{L}_{att} + \lambda\mathcal{L}_{ctc}
  \label{eq1}
 \vspace{-0.1cm}
\end{equation}
where the task weight $\lambda$ is empirically set as 0.2 during training and fixed throughout the experiments of this paper.


\vspace{-0.2cm}
\subsection{Conformer LHUC Speaker Adaptation}
\vspace{-0.1cm}
The key idea of learning hidden unit contributions (LHUC) \cite{Swietojanski2014LearningHU,swietojanski2016learning,xie2021bayesian} and the related parameterised adaptive activation functions \cite{Zhang2016DNNSA,Huang2017BayesianUB} is to modify the amplitudes of hidden unit activations of hybrid DNN, or Conformer models as considered in this paper, for each speaker using a speaker dependent (SD) transformation. This can be parameterized by using activation output scaling vectors. Let ${\bm{r}}^{l,s}$ denote the SD parameters for speaker $s$ in the $l$-th hidden layer, the hidden layer output is
\vspace{-0.1cm}
\begin{equation}
{\bm{h}}^{l,s}=\xi(\bm{r}^{l,s})\odot {\bm{h}}^{l},
\vspace{-0.1cm}
\end{equation}
where $\odot$ denotes the Hadamard product operation, ${\bm{h}}^{l}$ is the hidden vector after a non-linear activation function in the $l$-th layer, and $\xi(\bm{r}^{l,s})$ is the scaling vector parametrized by ${\bm{r}}^{l,s}$. In this paper, $\xi(\cdot)$ is the element-wise 2Sigmoid$(\cdot)$ function with range $(0, 2)$. Given the adaptation data ${\cal{D}}^{s}=\{{\bm{X}}^{s},{\bm{Y}}^{s}\}$ for speaker $s$, ${\bm{X}}^{s}$ and ${\bm{Y}}^{s}$ stand for the acoustic features and the corresponding supervision token sequences, respectively. The estimation of SD parameters ${\bm{r}}^s$ can be obtained by minimizing the loss in Eq. (\ref{eq1}), which is given by
\begin{equation}
\setlength{\abovedisplayskip}{4pt} 
\setlength{\belowdisplayskip}{4pt}
\begin{aligned}
\hat{{\bm{r}}}^s =& \arg\min\limits_{{\bm{r}}^s}\{(1-\lambda)(-\log P_{att}({\bm Y}^{s}|{\bm X}^s, {\bm r}^s))  \\
&+\lambda (-\log P_{ctc}({\bm Y}^{s}|{\bm X}^s, {\bm r}^s)) \},
\end{aligned}
\end{equation}
where $P_{att}({\bm Y}^{s}|{\bm X}^s, {\bm r}^s)$ and $P_{ctc}({\bm Y}^{s}|{\bm X}^s, {\bm r}^s)$ are the attention-based and CTC-based likelihood probabilities. The supervisions ${\bm{Y}}^{s}$ that are normally unavailable during unsupervised test time adaptation can be obtained by initially decoding the corresponding utterances using a baseline speaker independent (SI) model, before serving as target labels in the adaptation stage. During cross-adaptation based system combination, such supervision needs to be produced by a different system in the initial recognition pass.

\vspace{-0.2cm}
\section{Hybrid CNN-TDNN System}
\vspace{-0.1cm}
The CNN-TDNN hybrid ASR system and Bayesian LHUC Speaker Adaptation method are presented in this section. 
\vspace{-0.2cm}
\subsection{CNN-TDNN ASR Architecture}
\vspace{-0.1cm}
TDNNs based hybrid HMM-DNN acoustic models in recent years defined state-of-the-art speech recognition performance over a wide range of tasks. In particular, the recently proposed lattice-free MMI trained factored TDNN (TDNN-F) systems \cite{gulanticonformer2020} benefits from a compact model structure featuring low-rank weight matrix factorization while remain highly competitive performance wise against end-to-end approaches to date.

TDNNs can be considered as a special form of one-dimensional CNNs when parameters are tied across different time steps. The bottom layers of TDNNs are designed to learn a narrower temporal context span, while the higher layers to learn wider, longer range temporal contexts. To further reduce the risk of overfitting to limited training data and the number of parameters, a factored TDNN model structure was proposed in \cite{povey2018semi}, which compresses the weight matrix by using semi-orthogonal low-rank matrix factorization. The baseline hybrid CNN-TDNN system architecture follows the recipe\footnote{egs/swbd/s5c/local/chain/run\_cnn\_tdnn\_1a.sh} of the Kaldi tookit \cite{povey2011kaldi} and its more detailed description is in \cite{xie2021bayesian}\footnote{The baseline BLHUC results used by this paper are introduced in the appendix of arxiv version}. The network consisted of 15 context-splicing layers. A 160-dimensional factored linear projection was employed prior to affine transformation in each context-splicing layer other than the first one. ReLU activation is used in each context-splicing layer, followed by batch normalization and dropout. The hidden layer input prior to context splicing was scaled and added to the hidden layer output by a skip connection. After two linear projection layers and batch normalization, the output layer generated the probabilities tied HMM states. The CNN-TDNN architecture further employed 2-dimensional convolutional layers as the first 6 layers, where the two dimensions of the feature maps were the “height” and “time” axes. An example CNN-TDNN hybrid system is shown in Figure~\ref{fig:architecture} (grey box, bottom). 
\vspace{-0.6cm}
\subsection{CNN-TDNN Bayesian LHUC Speaker Adaptation}
\vspace{-0.1cm}
Limited speaker level adaptation data often leads to sparsity issue and modelling uncertainty for conventional LHUC adaptation using fixed value parameter estimation. To this end, the solution adopted in this paper use Bayesian learning by modelling SD LHUC parameters’ uncertainty in CNN-TDNN system adaptation. The following predictive distribution is utilized in Bayesian LHUC adaptation. The prediction over the test utterance ${\tilde{\bm X}}^s$ is given by
\vspace{-0.3cm}
\begin{equation}
p(\tilde{\bm Y}^s|\tilde{\bm X}^s,{\cal D}^s)=\int{p(\tilde{\bm{Y}}^{s}|\tilde{\bm{X}}^{s},{\bm r}^s)p({\bm r}^s|{\cal D}^s)d{\bm r}^s},
\label{eq:posterior}
\vspace{-0.3cm}
\end{equation}
\setlength{\abovedisplayskip}{4pt} 
\setlength{\belowdisplayskip}{4pt}
where ${\tilde{\bm Y}}^s$ is the predicted token sequence and $p({\bm r}^s|{\cal D}^s)$ is the SD parameters posterior distribution. The true posterior distribution can be approximated with a variational distribution $q({\bm r}^s)$ by minimising the following bound of cross entropy marginalisation over the adaptation data set ${\cal D}^s$,
\begin{align}
&-\log{p({\bm Y}^{s}|{\bm X}^s)}=-\log\int{p({\bm Y}^{s}, {\bm r}^s|{\bm X}^s)d{\bm r}^s} \nonumber \\
\leq& -\int{q({\bm r}^s)\log p({\bm Y}^{s}|{\bm X}^s, {\bm r}^s))d{\bm r}^s}+KL(q({\bm r}^s)||p({\bm r}^s))\nonumber\\
\vspace{-0.2cm}
\triangleq&\quad{\cal L}_1 + {\cal L}_2,
\end{align}
where $p({\bm r}^s)$ is the prior distribution of SD parameters, and KL$(\cdot)$ is the KL divergence. For simplicity, both $q({\bm r}^s)={\cal N}({\bm \mu}, {\bm \sigma}^2)$ and $p({\bm r}^s)={\cal N}({\bm \mu}_r, {\bm \sigma}_r^2)$ are assumed to be normal distributions. The first term ${\cal L}_1$ is approximated with Monte Carlo sampling method, which is given by
\vspace{-0.2cm}
\begin{align}
{\cal L}_1\approx \frac{1}{N}\sum\limits_{k=1}^{N}{\log p({\bm Y}^{s}|{\bm X}^s, {\bm \mu}+{\bm \sigma}\odot {\bm \epsilon}_k)},
\label{eq:loss1}
\vspace{-0.3cm}
\end{align}
where ${\bm \epsilon}_k$ is the $k$-th Monte Carlo sampling value drawn from the standard normal distribution. The KL divergence ${\cal L}_2$ can be explicitly calculated as
\vspace{-0.1cm}
\begin{align}
\mathcal{L}_2=\frac{1}{2}\sum_{i}{(\frac{\sigma_i^2+(\mu_i-\mu_{r,i})^2}{\sigma_{r,i}^2}+2\log\frac{\sigma_{r,i}}{\sigma_{i}}-1}),
\vspace{-0.3cm}
\end{align}
where $\{\mu_{r,i},\sigma_{r,i}\}$ and $\{\mu_i,\sigma_i\}$ are the $i$-th elements of vectors $\{{\bm \mu}_r,{\bm \sigma}_r\}$ and $\{{\bm \mu},{\bm \sigma}\}$, respectively.

\textbf{Implementation details} over several crucial settings are: 1) Based on empirical evaluation and following the configuration used in \cite{xie2021bayesian}, the LHUC transforms are applied to the hidden output of the non-convolutional hidden layer outs (starting from layer 7 onwards). 2) The LHUC prior distribution is modelled by a standard normal distribution ${\cal N}(\bf 0, 1)$. 3) In Bayesian learning, only one parameter sample is drawn in (\ref{eq:loss1}) during adaptation to ensure that the computational cost is comparable to that of standard LHUC adaptation. 4) The predictive inference integral in (\ref{eq:posterior}) is efficiently approximated by the expectation of the posterior distribution as $p(\tilde{\bm{Y}}^{s}|\tilde{\bm{X}}^{s},{{\mathbb{E}}[{\bm r}^s|{\cal D}^s]})$.

A similar Bayesian adaptation approach may also be considered for Conformer systems \cite{deng2022confidence}. However, in order to create sufficient cross system diversity, Bayesian adaptation is only applied to hybrid CNN-TDNN systems in this paper. 

\section{System Combination}
Two-pass rescoring and cross adaptation based system combination are presented in this section. 
\vspace{-0.2cm}
\subsection{Two-pass Rescoring}
\vspace{-0.1cm}
Since the successful application of multi-pass decoding based system combination \cite{woodland19951994, hain2005automatic, hain2003automatic} to traditional HMM based ASR system, their efficacy has been found to crucially depend on the quality of first decoding pass that serves to produce the initial recognition outputs. As the first pass decoding outputs serve as the constraint of the following recognition stages, it is important for the system used in the initial recognition pass to produce compact lattice based hypotheses. To this end, efficient decoding algorithms were extensively researched in the context of HMM and hybrid HMM-DNN systems \cite{odell1994one, ney1999dynamic, ortmanns2000look, rybach2013lexical, mohri2002weighted}.

In contrast, the attention-based encoder-decoder model architecture and auto-regressive decoder component design utilizing the full history context used in Conformer, and other related E2E models \cite{Prabhavalkar2021LessIM}, leads to their difficulty in deploying effective beam search in lattice based decoding. Hence, hybrid CNN-TDNNs are used in the first pass beam search decoding of this paper to produce an initial N-best outputs. These are then rescored using LHUC speaker adapted Conformer \cite{deng2022confidence} via a 2-way cross system sequence level score interpolation. 



Given the acoustic features ${\bm X}$ and its corresponding N-best recognition hypotheses in $\{{\bm Y_1}, {\bm Y_2}, \cdots, {\bm Y_n}\}$. Let ${\bm s}^{cfm}=[s_{1}^{cfm}, s_{2}^{cfm}, \cdots, s_{n}^{cfm}]^{\top}$ and ${\bm s}^{tdnn}=[s_{1}^{tdnn}, s_{2}^{tdnn}, \cdots, s_{n}^{tdnn}]^{\top}$ denote the N-best score vectors respectively produced by the Conformer (cfm) and CNN-TDNN (tdnn) systems. 
The system combination 1-best output via two-pass Conformer rescoring is then obtained by 
\vspace{-0.2cm}
\begin{align}
{\hat{\bm Y}_{best}} = \arg\min\limits_{i}\left\{\beta {s}_{i}^{cfm} + (1-\beta) {s}_{i}^{tdnn} \right\}
\vspace{-0.2cm}
\end{align}
where $\beta \in [0,1]$ is the weight assigned to the Conformer system. Example two-pass rescoring based system combination is shown in Figure~\ref{fig:architecture} (purple dotted line).


\begin{table*}[htbp]
\vspace{-1.5cm}
\setlength{\abovecaptionskip}{-0cm}
\centering
\caption{Performance (WER\%) of two-pass decoding and cross adaptation based system combination of Conformer (cfm) and CNN-TDNN systems evaluated on the Hub5’00, RT02 and RT03 data sets. "$X \rightarrow Y$" stands for N-best (N=100) outputs of system "X" rescored by "Y" via score interpolation of Section 4.1; "$X \Longrightarrow Y$" stands for system "Y" being cross adapted to 1-best outputs of "X" in combination as Section 4.2. $\dagger$ denotes a statistically significant WER difference over the baseline systems (sys. 1, 2, 3)}
\label{tab:table1}
\resizebox{2.1\columnwidth}{!}{
\begin{tabular}{c|l|c|cc|ccc|ccc|cccc} 
\hline\hline 
    \multirow{2}{*}{ID} & 
    \multirow{2}{*}{System}  &
    \multirow{2}{*}{\tabincell{c}{System\\Combination}} &
    \multicolumn{2}{c|}{System Weight} &
\multicolumn{3}{c|}{Hub5' 00} &
\multicolumn{3}{c|}{Rt03S} &
\multicolumn{4}{c}{Rt02} \\
\cline{4-15}
&  & & CFM & TDNN & SWB1 & CHM & Avg  & SWB2 & FSH & Avg & SWB3 & SWB4 & SWB5 & Avg   \\
\hline\hline

1 & LF-MMI CNN-TDNN + BLHUC-SAT \cite{xie2021bayesian} & - &  \multicolumn{2}{c|}{-} &6.5 & 12.2 & 9.4  &  12.9 & 7.7 & 10.4 & 7.6 & 9.8 & 12.0 & 9.9 \\
2 & ESPnet Conformer & - &  \multicolumn{2}{c|}{-} &7.3 & 14.9 & 11.1 & 16.4 & 10.4 & 13.5 & 8.7 & 12.9 & 15.6 & 12.6 \\
3 & ESPnet Conformer + LHUC-SAT \cite{deng2022confidence}  & - &  \multicolumn{2}{c|}{-} &7.2 & 13.8 & 10.5 & 15.7 & 10.0 & 12.9 & 8.6 & 12.3 & 14.7 & 12.0   \\
\hline\hline

4 & \multirow{10}*{Sys. 1 $\rightarrow$ Sys. 2} &  \multirow{10}*{\tabincell{c}{Two-pass\\rescoring}} &  \multicolumn{2}{c|}{0.1 / 0.9} & 6.1 & 11.6 & 8.9 & 12.2 & 7.3 & 9.9 & 7.0 & 9.3 &  11.4 & 9.4   \\
5 & & &  \multicolumn{2}{c|}{0.2 / 0.8} & 5.9 & 11.3 & 8.7 & 12.0 & 7.1 & 9.6 & 6.9 & 9.2 & 11.3& 9.3   \\
6 & & &  \multicolumn{2}{c|}{0.3 / 0.7} & ${\textbf{5.9}^{\dagger}}$ & ${\textbf{11.3}^{\dagger}}$ & ${\textbf{8.6}^{\dagger}}$ & ${\textbf{11.9}^{\dagger}}$ & ${\textbf{7.0}^{\dagger}}$ & ${\textbf{9.6}^{\dagger}}$ & ${\textbf{6.9}^{\dagger}}$ & ${\textbf{9.3}^{\dagger}}$ & ${\textbf{11.1}^{\dagger}}$  & ${\textbf{9.3}^{\dagger}}$\\
7 &  & &  \multicolumn{2}{c|}{0.4 / 0.6} & 6.0 & 11.4 & 8.7 & 12.0 & 7.2 & 9.7 & 7.0 & 9.3 & 11.1 & 9.3   \\
8 &  & &  \multicolumn{2}{c|}{0.5 / 0.5} & 6.2 & 11.5 & 8.9 & 12.3 & 7.4 & 9.9 & 7.2 & 9.5 & 11.2 & 9.4   \\
9 &  & &  \multicolumn{2}{c|}{0.6 / 0.4} & 6.3 & 11.9 & 9.1 & 12.6 & 7.6 & 10.2 & 7.3 & 9.6 & 11.3 & 9.6   \\
10 &  & &  \multicolumn{2}{c|}{0.7 / 0.3} & 6.4 & 12.0 & 9.2 & 12.7 & 7.7 & 10.3 & 7.5 & 9.9 & 11.6 & 9.8  \\
11 &  & &  \multicolumn{2}{c|}{0.8 / 0.2} & 6.5 & 12.2 & 9.4 & 12.9 & 7.8 & 10.5 & 7.6 & 10.0 & 11.8 & 9.9   \\
12 &  & &  \multicolumn{2}{c|}{0.9 / 0.1} & 6.5 & 12.4 & 9.5 & 13.1 & 8.0 & 10.7 & 7.7 & 10.1 & 11.9 & 10.1   \\
13 &  & &  \multicolumn{2}{c|}{1.0 / 0.0}& 6.6 & 12.6 & 9.6 & 13.2 & 8.1 & 10.8 & 7.7 & 10.2 & 12.0 & 10.1   \\
\hline
14 & \multirow{10}*{Sys. 1 $\rightarrow$ Sys. 3} &\multirow{10}*{\tabincell{c}{Two-pass\\rescoring}} & \multicolumn{2}{c|}{0.1 / 0.9} & 6.2 & 11.9 & 9.1 & 12.5 & 7.5 & 10.1 & 7.2 & 9.5 & 11.6 & 9.6\\

15 & & & \multicolumn{2}{c|}{0.2 / 0.8} &
${\textbf{6.1}^{\dagger}}$ & ${\textbf{11.9}^{\dagger}}$ & ${\textbf{9.0}^{\dagger}}$ & ${\textbf{12.4}^{\dagger}}$ & ${\textbf{7.4}^{\dagger}}$ & ${\textbf{10.0}^{\dagger}}$ & ${\textbf{7.0}^{\dagger}}$ & ${\textbf{9.4}^{\dagger}}$ & ${\textbf{11.5}^{\dagger}}$  & ${\textbf{9.5}^{\dagger}}$
\\
16 & & & \multicolumn{2}{c|}{0.3 / 0.7} & 6.2 & 11.9 & 9.1 & 12.5 & 7.5 & 10.1 & 7.1 & 9.5 & 11.6 & 9.6\\
17 & & & \multicolumn{2}{c|}{0.4 / 0.6} & 6.3 & 12.1 & 9.2 & 12.6 & 7.6 & 10.2 & 7.2 & 9.6 & 11.7 & 9.7\\
18 & & & \multicolumn{2}{c|}{0.5 / 0.5} & 6.5 & 12.3 & 9.4 & 12.9 & 7.8 & 10.4 & 7.3 & 9.8 & 11.9 & 9.8\\
19 & & & \multicolumn{2}{c|}{0.6 / 0.4} & 6.7 & 12.6 & 9.7 & 13.2 & 8.0 & 10.7 & 7.5 & 10.0 & 12.2 & 10.1\\
20 & & & \multicolumn{2}{c|}{0.7 / 0.3} & 6.9 & 12.9 & 9.9 & 13.4 & 8.3 & 10.9 & 7.7 & 10.2 & 12.5 & 10.3 \\
21 & & & \multicolumn{2}{c|}{0.8 / 0.2} & 7.0 & 13.2 & 10.1 & 13.7 & 8.5 & 11.2& 7.9 & 10.5 & 12.8 & 10.5\\
22 & & & \multicolumn{2}{c|}{0.9 / 0.1} & 7.1 & 13.3 & 10.3 & 13.9 & 8.8 & 11.4 & 8.1 & 10.6 & 13.0 & 10.8\\
23 & & & \multicolumn{2}{c|}{1.0 / 0.0} & 7.3 & 13.5 & 10.4 & 14.1 & 9.0 & 11.6 & 8.2 & 10.8 & 13.3 & 10.9\\
\hline\hline
24 & Sys. 1 $\Longrightarrow$ Sys. 3 & \multirow{2}*{\tabincell{c}{Cross\\adaptation}} &  \multicolumn{2}{c|}{-}  & ${6.2}^{\dagger}$ & ${13.4}^{\dagger}$ & ${10.3}^{\dagger}$ &${15.1}^{\dagger}$ & ${9.4}^{\dagger}$ & ${12.4}^{\dagger}$ & ${8.3}^{\dagger}$ & ${11.0}^{\dagger}$ & ${12.7}^{\dagger}$ & ${10.8}^{\dagger}$ \\
25 & Sys. 3 $\Longrightarrow$ Sys. 1 & &  \multicolumn{2}{c|}{-}  & 
${\textbf{6.2}^{\dagger}}$ & ${\textbf{12.2}^{\dagger}}$ & ${\textbf{9.2}^{\dagger}}$ & ${\textbf{12.0}^{\dagger}}$ & ${\textbf{7.2}^{\dagger}}$ & ${\textbf{9.7}^{\dagger}}$ & ${\textbf{7.1}^{\dagger}}$ & ${\textbf{9.5}^{\dagger}}$ & ${\textbf{11.1}^{\dagger}}$  & ${\textbf{9.4}^{\dagger}}$
\\
\hline\hline
\end{tabular}
}
\vspace{-0.5cm}
\end{table*}
\vspace{-0.2cm}
\subsection{Cross Adaptation}
\vspace{-0.1cm}
When there is a large variation of recognition error rate performance among component systems, for example, the significant WER differences of 10\% to 20\% relative found between the standalone Conformer and CNN-TDNN systems of the top section of Table~\ref{tab:table1} (sys. 1, 2, 3), 
cross adaptation provides an alternative to multi-pass decoding based system combination. The aim of cross system adaptation is to shift the underlying decision boundary of one component system by learning the useful but distinct characteristics from another system whose recognition outputs serve as the adaptation supervision. In this paper, two cross-adaptation based combination configurations are considered. The CNN-TDNN system is cross adapted to the 1-best output of the Conformer system or vice versa. These are shown in Figure 1 using the blue and green dotted lines respectively.


\vspace{-0.2cm}
\section{Experiment}
In this section, the performance of two-pass rescoring and cross adaptation combination of CNN-TDNN and Conformer systems are evaluated on the 300-hr Switchboard task. 
\vspace{-0.2cm}
\subsection{Experiment Setup}
\vspace{-0.1cm}
The Switchboard-1 corpus\cite{godfrey1992switchboard} with 286-hr speech collected from 4804 speakers is used for training. The NIST Hub5’00, RT02 and RT03s evaluation sets containing 80, 120 and 144 speakers, and 3.8, 6.4 and 6.2 hours of speech respectively were used. In addition to the E2E Conformer system description of Section 2.2, 80-dim Mel-filter bank plus 3-dim pitch parameters were used as input features. The ESPNET recipe \cite{guo2021recent} configured Conformer model contains 12 encoder and 6 decoder layers, where each layer is configured with 4-head attention of 256-dim, and 2048 feed forward hidden nodes. Byte-pair-encoding (BPE) tokens served as the decoder outputs. The convolution subsampling module contains two 2-D convolutional layers with stride 2. SpecAugment \cite{park2019specaugment} and dropout (rate set as 0.1) were used in training, together with model averaging performed over the last ten epochs. 

Further to the description of the hybrid LF-MMI trained CNN-TDNN systems in Section 3.1, 40-dim Mel filter-bank features were transformed by linear discriminative analysis to produce the input features. The splicing indices $\{-1, 0, 1\}$ and $\{-3, 0, 3\}$ were employed in the first 4 and the last 10 context-splicing layers respectively. Speed perturbation, SpecAugment and RNNLM rescoring were also used. More detailed description can be found in \cite{xie2021bayesian}. Matched pairs sentence-segment word error (MAPSSWE) based statistical significance test was performed at a significance level $\alpha=0.05$ for all systems. 

\begin{table}[htbp]
\centering
\vspace{-0.2cm}
\normalsize
\setlength{\abovecaptionskip}{-0cm}
\caption{Performance (WER\%) contrasts of the two-pass decoding based system combination between Conformer and CNN-TDNN systems (system weights 0.3:0.7) w.r.t. N-best list depth.}
\label{tab:table2}
\resizebox{\columnwidth}{!}{
\begin{tabular}{c|l|c|cc|ccc|ccc|cccc} 
\hline\hline 
    \multirow{2}{*}{ID} & 
    \multirow{2}{*}{System}  &
    \multirow{2}{*}{N-best} &
    \multicolumn{2}{c|}{System Weight} &
\multicolumn{3}{c|}{Hub5' 00} &
\multicolumn{3}{c|}{Rt03S} &
\multicolumn{4}{c}{Rt02} \\
\cline{4-15}
&  & & CFM & TDNN & SWB1 & CHM & Avg  & SWB2 & FSH & Avg & SWB3 & SWB4 & SWB5 & Avg   \\
\hline\hline

1 & \multirow{4}*{\tabincell{c}{Multi-pass\\rescoring}} & 10 &  \multicolumn{2}{c|}{0.3 / 0.7} & 6.0 & 11.5 & 8.7 & 12.3 & 7.2 & 9.8 & 7.0 & 9.4 & 11.6 & 9.5 \\
2 & & 25 &  \multicolumn{2}{c|}{0.3 / 0.7} & 5.9 & 11.3 & 8.6 & 12.0 & 7.1 & 9.7 & 7.0 & 9.3 & 11.3 & 9.4 \\
3 &  & 50 &  \multicolumn{2}{c|}{0.3 / 0.7} & 6.0 & 11.2 & 8.6 &  11.9 & 7.0 & 9.6 & 6.9 & 9.3 & 11.2 & 9.3\\
4 &  & 100 &  \multicolumn{2}{c|}{0.3 / 0.7} &5.9 & 11.3 & 8.6  & 11.9 & 7.0 & 9.6 & 6.9 & 9.3 & 11.1 & 9.3  \\

\hline\hline
\end{tabular}
}
\end{table}

\vspace{-0.6cm}
\subsection{Performance of System Combination}
\vspace{-0.1cm}
The performance of various two-pass decoding and cross adaptation based system combination of Conformer and CNN-TDNN systems evaluated on the Hub5’00, RT02 and RT03 data are shown in Table 1 (sys. 4 to 25). Several trends can be found. First, using the SI or LHUC-SAT speaker adapted Conformer systems (sys. 2 or 3) to rescore the 100-best outputs of the CNN-TDNN system (sys. 1) in a second pass decoding, irrespective of the choice of system weighs, the two-pass decoding combined system outputs (sys. 4 to 25) consistently outperformed the stand alone SI or speaker adapted Conformer systems (sys. 2 or 3). Second, for both groups of two-pass decoding experiments (sys. 4-13, sys. 14-23), the best combination performance was obtained using a Conformer system weighting of 0.2-0.3 (sys. 6, 15). The best two-pass combined system outperformed the stand alone baseline SI Conformer system by statistically significant WER reductions of 2.5\%, 3.9\% and 3.3\% absolute (22.5\%, 28.9\% and 26.2\% relative) on the {\bf Hub5’00}, {\bf RT02} and {\bf RT03} test sets respectively. 
Further ablation studies in Table 3 analyse the performance of two-pass decoding based system combination of Conformer and CNN-TDNN systems (system weights 0.3:0.7) w.r.t. the N-best list depth, where much shallower N-best lists (e.g. N=25 or 10) can be used without significant WER increase.
Finally, the cross-adaption combined systems, whether using the CNN-TDNN or Conformer system’s 1-best outputs as supervision (sys. 24 or 25), consistently outperformed the stand alone SI and LHUC-SAT Conformer systems (sys. 2, 3). The best cross adaptation combined system involving cross adapting the CNN-TDNN system to the 1-best outputs of the Conformer LHUC-SAT system, outperformed the stand alone SI Conformer system by statistically significant WER reductions of 1.9\%, 3.8\% and 3.2\% absolute (17.1\%, 28.1\% and 25.4\% relative) on the three test sets. 

The performance of the best two-pass or cross adaptation combined systems (sys. 6, 25, Table~\ref{tab:table1}) are further contrasted in Table~\ref{tab:table3} with the state-of-the-art performance on the same task obtained using most recent hybrid and E2E systems reported in the literature to demonstrate their competitiveness. 

\begin{table}[htbp]
\vspace{-0.2cm}
\setlength{\abovecaptionskip}{-0cm}
\centering
\caption{Performance (WER\%) contrasts between best two combined systems and other SOTA results on Hub5'00 and RT03s}
\label{tab:table3}
\resizebox{\columnwidth}{!}{
\begin{tabular}{c|l|l|ccc|ccc} %
	\hline\hline
    \multirow{2}{*}{ID} & 
    \multirow{2}{*}{System}  &
    \multirow{2}{*}{\# Param.} &
    \multicolumn{3}{c|}{Hub5'00 } & 
    \multicolumn{3}{c}{Rt03S} 
    \\   \cline{4-9}
	& &  & SWB1 & CHM & Avg & SWB2 & FSH & Avg \\ \hline\hline
	1 & RWTH-2019 Hybrid\cite{Kitza2019CumulativeAF} & - & 6.7 &  13.5 & 10.2  & - & - & - \\ \hline
	2 & Google-2019 LAS\cite{Park2019SpecAugmentAS} & -  & 6.8 & 14.1 & (10.5) &- &- &- \\ \hline
	3 & \multirow{3}{*}{\tabincell{l}{IBM-2020 AED\cite{tuske2020single}}}  & 29M & 7.4 & 14.6 & (11.0)  & - & - & - \\
	4 &   & 75M  & 6.8 & 13.4 & (10.1)&  - & - & - \\
	5 &   & 280M & 6.4 & 12.5 & 9.5 & 14.8 & 8.4 & (11.7) \\ \hline
	6 & IBM-2021 CFM-AED \cite{Tuske2021OnTL} & 68M  & 5.5 & 11.2 & (8.4) & 12.6 & 7.0 & (9.9) \\ \hline
	7 & Salesforce-2020 \cite{wang2020investigation} & -  & 6.3 & 13.3 & (9.8) & - &- & 11.4 \\ \hline
	8 & Espnet-2021 CFM\cite{guo2021recent} & -  & 7.3 & 14.9 & 11.1  & 16.4 & 10.4 & 13.5 \\\hline
	9 & CUHK-2022 CFM-LHUC\cite{deng2022confidence} & 46M  & 7.2 & 13.8 & 10.5  & 15.7 & 10.0 & 12.9 
	\\ \hline
	10 & CUHK-2021 BLHUC \cite{xie2021bayesian} & 15.2M  & 6.5 & 12.2 & 9.4 & 12.8 & 7.6 & 10.3 \\
	\hline
	11 & $\textbf{Cross adaptation (Sys.25, Table~\ref{tab:table1}) (ours)} $& 61M & 6.2 & 12.2 & 9.2  & 12.0 & 7.2 & 9.7 \\
	12 & $\textbf{Two-pass rescoring (Sys.6, Table~\ref{tab:table1}) (ours)}$ & 61M & 5.9 & 11.3 & 8.6  & 11.9 & 7.0 & 9.6  \\\hline \hline
\end{tabular} 
 }
\vspace{-0.5cm}
\end{table}

\vspace{-0.2cm}
\section{Conclusions}
\vspace{-0.1cm}
In this paper, we presented the first use of multi-pass rescoring and cross adaptation based system combination methods to integrate state-of-the-art speaker adapted hybrid CNN-TDNN and Conformer ASR systems. Experiments on the 300-hour Switchboard task suggest that the combined systems derived using either of the two system combination approaches outperformed the individual systems by up to 3.9\% absolute (28.9\% relative) reduction in WER over the stand alone Conformer system on the NIST {\bf Hub5’00}, {\bf Rt03} and {\bf Rt02} evaluation data. Future researches will focus on improving cross system complementarity and diversity for combination. 

\vspace{-0.2cm}
\section{Acknowledgements}
\vspace{-0.1cm}
This research is supported by Hong Kong RGC GRF grant No. 14200021, 14200218, 14200220, Innovation \& Technology Fund grant No. ITS/254/19 and ITS/218/21.

\bibliographystyle{IEEEtran}

\bibliography{mybib}

\begin{thebibliography}{10}
\providecommand{\url}[1]{#1}
\csname url@samestyle\endcsname
\providecommand{\newblock}{\relax}
\providecommand{\bibinfo}[2]{#2}
\providecommand{\BIBentrySTDinterwordspacing}{\spaceskip=0pt\relax}
\providecommand{\BIBentryALTinterwordstretchfactor}{4}
\providecommand{\BIBentryALTinterwordspacing}{\spaceskip=\fontdimen2\font plus
\BIBentryALTinterwordstretchfactor\fontdimen3\font minus
  \fontdimen4\font\relax}
\providecommand{\BIBforeignlanguage}[2]{{%
\expandafter\ifx\csname l@#1\endcsname\relax
\typeout{** WARNING: IEEEtran.bst: No hyphenation pattern has been}%
\typeout{** loaded for the language `#1'. Using the pattern for}%
\typeout{** the default language instead.}%
\else
\language=\csname l@#1\endcsname
\fi
#2}}
\providecommand{\BIBdecl}{\relax}
\BIBdecl

\bibitem{vesely2013sequence}
K.~Vesel{\`y} \emph{et~al.}, ``Sequence-discriminative training of deep neural
  networks.'' in \emph{INTERSPEECH}, 2013.

\bibitem{povey2018semi}
D.~Povey \emph{et~al.}, ``Semi-orthogonal low-rank matrix factorization for
  deep neural networks.'' in \emph{INTERSPEECH}, 2018.

\bibitem{abdel2012applying}
O.~Abdel-Hamid \emph{et~al.}, ``Applying convolutional neural networks concepts
  to hybrid nn-hmm model for speech recognition,'' in \emph{ICASSP}, 2012.

\bibitem{medsker2001recurrent}
L.~R. Medsker \emph{et~al.}, ``Recurrent neural networks,'' \emph{Design and
  Applications}, 2001.

\bibitem{graves2013speech}
A.~Graves \emph{et~al.}, ``Speech recognition with deep recurrent neural
  networks,'' in \emph{ICASSP}, 2013.

\bibitem{sak2014long}
H.~Sak \emph{et~al.}, ``Long short-term memory based recurrent neural network
  architectures for large vocabulary speech recognition,'' \emph{arXiv preprint
  arXiv:1402.1128}, 2014.

\bibitem{peddinti2015time}
V.~Peddinti \emph{et~al.}, ``A time delay neural network architecture for
  efficient modeling of long temporal contexts,'' in \emph{ISCA}, 2015.

\bibitem{povey2016purely}
D.~Povey \emph{et~al.}, ``Purely sequence-trained neural networks for asr based
  on lattice-free mmi.'' in \emph{INTERSPEECH}, 2016.

\bibitem{chan2016listen}
W.~Chan \emph{et~al.}, ``Listen, attend and spell: A neural network for large
  vocabulary conversational speech recognition,'' in \emph{ICASSP}, 2016.

\bibitem{graves2012sequence}
A.~Graves, ``Sequence transduction with recurrent neural networks,''
  \emph{arXiv preprint arXiv:1211.3711}, 2012.

\bibitem{vaswani2017attention}
A.~Vaswani \emph{et~al.}, ``Attention is all you need,'' \emph{Advances in
  neural information processing systems}, 2017.

\bibitem{dong2018speech}
L.~Dong \emph{et~al.}, ``Speech-transformer: a no-recurrence
  sequence-to-sequence model for speech recognition,'' in \emph{ICASSP}, 2018.

\bibitem{gulati2020conformer}
A.~Gulati \emph{et~al.}, ``Conformer: Convolution-augmented transformer for
  speech recognition,'' \emph{arXiv preprint arXiv:2005.08100}, 2020.

\bibitem{guo2021recent}
P.~Guo \emph{et~al.}, ``Recent developments on espnet toolkit boosted by
  conformer,'' in \emph{ICASSP}, 2021.

\bibitem{karita2019comparative}
S.~Karita \emph{et~al.}, ``A comparative study on transformer vs rnn in speech
  applications,'' in \emph{ASRU}, 2019.

\bibitem{woodland2004superears}
P.~Woodland \emph{et~al.}, ``Superears: Multi-site broadcast news system,'' in
  \emph{Rich Transcription (RT-04F) Workshop}, 2004.

\bibitem{schwartz2004speech}
R.~Schwartz \emph{et~al.}, ``Speech recognition in multiple languages and
  domains: the 2003 bbn/limsi ears system,'' in \emph{ICASSP}, 2004.

\bibitem{lei2009development}
X.~Lei \emph{et~al.}, ``Development of the 2008 sri mandarin speech-to-text
  system for broadcast news and conversation,'' in \emph{ISCA}, 2009.

\bibitem{chu20102009}
S.~M. Chu \emph{et~al.}, ``The 2009 ibm gale mandarin broadcast transcription
  system,'' in \emph{ICASSP}, 2010.

\bibitem{liu2013language}
X.~Liu \emph{et~al.}, ``Language model cross adaptation for lvcsr system
  combination,'' \emph{CSL}, 2013.

\bibitem{lamel2011improved}
L.~Lamel \emph{et~al.}, ``Improved models for mandarin speech-to-text
  transcription,'' in \emph{ICASSP}, 2011.

\bibitem{woodland19951994}
P.~C. Woodland \emph{et~al.}, ``The 1994 htk large vocabulary speech
  recognition system,'' in \emph{ICASSP}, 1995.

\bibitem{hain2005automatic}
T.~Hain \emph{et~al.}, ``Automatic transcription of conversational telephone
  speech,'' \emph{IEEE/ACM TSALP}, 2005.

\bibitem{hain2003automatic}
T.~Hain \emph{et~al.}, ``Automatic transcription of conversational telephone
  speech-development of the cu-htk 2002 system,'' in \emph{ICASSP}, 2003.

\bibitem{peskin1999improvements}
B.~Peskin \emph{et~al.}, ``Improvements in recognition of conversational
  telephone speech,'' in \emph{ICASSP}, 1999.

\bibitem{prasad20052004}
R.~Prasad \emph{et~al.}, ``The 2004 bbn/limsi 20xrt english conversational
  telephone speech recognition system,'' in \emph{ECSCT}, 2005.

\bibitem{fiscus1997post}
J.~G. Fiscus, ``A post-processing system to yield reduced word error rates:
  Recognizer output voting error reduction (rover),'' in \emph{ASRUP Workshop},
  1997.

\bibitem{evermann2000posterior}
G.~Evermann \emph{et~al.}, ``Posterior probability decoding, confidence
  estimation and system combination,'' in \emph{Proc. Speech Transcription
  Workshop}, 2000.

\bibitem{watanabe2017hybrid}
S.~Watanabe \emph{et~al.}, ``Hybrid ctc/attention architecture for end-to-end
  speech recognition,'' \emph{Selected Topics in Signal Processing}, 2017.

\bibitem{sainath2019two}
T.~N. Sainath \emph{et~al.}, ``Two-pass end-to-end speech recognition,''
  \emph{arXiv preprint arXiv:1908.10992}, 2019.

\bibitem{li2019integrating}
Q.~Li \emph{et~al.}, ``Integrating source-channel and attention-based
  sequence-to-sequence models for speech recognition,'' in \emph{ASRU}, 2019.

\bibitem{park2019specaugment}
D.~S. Park \emph{et~al.}, ``Specaugment: A simple data augmentation method for
  automatic speech recognition,'' \emph{arXiv preprint arXiv:1904.08779}, 2019.

\bibitem{xie2019blhuc}
X.~Xie \emph{et~al.}, ``Blhuc: Bayesian learning of hidden unit contributions
  for deep neural network speaker adaptation,'' in \emph{ICASSP}, 2019.

\bibitem{wong2020combination}
J.~H. Wong \emph{et~al.}, ``Combination of end-to-end and hybrid models for
  speech recognition.'' in \emph{INTERSPEECH}, 2020.

\bibitem{xie2021bayesian}
X.~Xie \emph{et~al.}, ``Bayesian learning for deep neural network adaptation,''
  \emph{IEEE/ACM TASLP}, 2021.

\bibitem{deng2022confidence}
J.~Deng \emph{et~al.}, ``Confidence score based conformer speaker adaptation
  for speech recognition,'' \emph{INTERSPEECH}, 2022.

\bibitem{dauphin2017language}
Y.~N. Dauphin \emph{et~al.}, ``Language modeling with gated convolutional
  networks,'' in \emph{ICML}, 2017.

\bibitem{Swietojanski2014LearningHU}
P.~Swietojanski \emph{et~al.}, ``Learning hidden unit contributions for
  unsupervised speaker adaptation of neural network acoustic models,''
  \emph{SLT Workshop}, 2014.

\bibitem{swietojanski2016learning}
P.~Swietojanski \emph{et~al.}, ``Learning hidden unit contributions for
  unsupervised acoustic model adaptation,'' \emph{IEEE/ACM TASLP}, 2016.

\bibitem{Zhang2016DNNSA}
C.~Zhang \emph{et~al.}, ``Dnn speaker adaptation using parameterised sigmoid
  and relu hidden activation functions,'' \emph{ICASSP}, 2016.

\bibitem{Huang2017BayesianUB}
Z.~Huang \emph{et~al.}, ``Bayesian unsupervised batch and online speaker
  adaptation of activation function parameters in deep models for automatic
  speech recognition,'' \emph{IEEE/ACM TASLP}, 2017.

\bibitem{gulanticonformer2020}
A.~Gulati \emph{et~al.}, ``Conformer: Convolution-augmented transformer for
  speech recognition,'' \emph{INTERSPEECH}, 2020.

\bibitem{povey2011kaldi}
D.~Povey \emph{et~al.}, ``The kaldi speech recognition toolkit,'' in
  \emph{Workshop on ASRU}, 2011.

\bibitem{odell1994one}
J.~J. Odell \emph{et~al.}, ``A one pass decoder design for large vocabulary
  recognition,'' in \emph{Human Language Technology}, 1994.

\bibitem{ney1999dynamic}
H.~Ney \emph{et~al.}, ``Dynamic programming search for continuous speech
  recognition,'' \emph{Signal Processing Magazine}, 1999.

\bibitem{ortmanns2000look}
S.~Ortmanns \emph{et~al.}, ``Look-ahead techniques for fast beam search,''
  \emph{CSL}, 2000.

\bibitem{rybach2013lexical}
D.~Rybach \emph{et~al.}, ``Lexical prefix tree and wfst: A comparison of two
  dynamic search concepts for lvcsr,'' \emph{IEEE/ACM TASLP}, 2013.

\bibitem{mohri2002weighted}
M.~Mohri \emph{et~al.}, ``Weighted finite-state transducers in speech
  recognition,'' \emph{CSL}, 2002.

\bibitem{Prabhavalkar2021LessIM}
R.~Prabhavalkar \emph{et~al.}, ``Less is more: Improved rnn-t decoding using
  limited label context and path merging,'' \emph{ICASSP}, 2021.

\bibitem{godfrey1992switchboard}
J.~J. Godfrey \emph{et~al.}, ``Switchboard: Telephone speech corpus for
  research and development,'' in \emph{ICASSP}, 1992.

\bibitem{Kitza2019CumulativeAF}
M.~Kitza \emph{et~al.}, ``Cumulative adaptation for blstm acoustic models,''
  \emph{INTERSPEECH}, 2019.

\bibitem{Park2019SpecAugmentAS}
D.~S. Park \emph{et~al.}, ``Specaugment: A simple data augmentation method for
  automatic speech recognition,'' in \emph{INTERSPEECH}, 2019.

\bibitem{tuske2020single}
Z.~T{\"u}ske \emph{et~al.}, ``Single headed attention based
  sequence-to-sequence model for state-of-the-art results on switchboard,''
  \emph{INTERSPEECH}, 2020.

\bibitem{Tuske2021OnTL}
Z.~Tuske \emph{et~al.}, ``On the limit of {English} conversational speech
  recognition,'' in \emph{INTERSPEECH}, 2021.

\bibitem{wang2020investigation}
W.~Wang \emph{et~al.}, ``An investigation of phone-based subword units for
  end-to-end speech recognition,'' \emph{INTERSPEECH}, 2020.

\end{thebibliography}


\end{document}